\documentclass[aps,prl,twocolumn,showpacs,superscriptaddress,tightenlines,preprintnumbers,sort&compass,amsmath,amssymb,floatfix]{revtex4}

\usepackage{comment}
\usepackage{amssymb}
\usepackage{amsmath}
\usepackage[dvipsnames,usenames]{color}
\usepackage{graphicx,graphics}
\usepackage{amssymb}

\newcommand\ket[1]{\ensuremath{|#1\rangle}}
\newcommand\bra[1]{\ensuremath{\langle#1|}}
\begin{document}
\title{Improving quantum entanglement through single-qubit operations}
\author{Xiang-Bin Wang}
\affiliation{Department of Physics and  State Key Laboratory of Low-Dimensional Quantum Physics, Tsinghua University,
Beijing 100084, China} \affiliation{Advanced Science Institute,
RIKEN, Wako-shi, Saitama, 351-0198, Japan }
\author{Zong-Wen Yu}
\affiliation{Department of Physics and  the Key Laboratory of Atomic
and Nanosciences, Ministry of Education, Tsinghua University,
Beijing 100084, China}
\author{Jia-Zhong Hu}
\affiliation{Department of Physics and  the Key Laboratory of Atomic
and Nanosciences, Ministry of Education, Tsinghua University,
Beijing 100084, China}
\author {Franco Nori}
\affiliation{Advanced Science Institute, RIKEN, Wako-shi, Saitama,
351-0198, Japan } \affiliation{Physics Department,The University of
Michigan, Ann Arbor, Michigan 48109-1040, USA }


\begin{abstract}
  We show that the entanglement of a $2\times 2$ bipartite state can be improved and maximized probabilistically through single-qubit operations only.
  An experiment is
  proposed and it is numerically simulated.
\end{abstract}

\pacs{03.65.Ud, 03.67.Ac}
\maketitle
\noindent{\em Introduction.---}
Quantum entanglement plays a central role in
 quantum information  and also in the
foundations of quantum physics. Thus, it has been extensively
studied (see, e.g.,~\cite{amico,horodecki,pv,Yutin,sci,nor}).
One important topic here is how to improve quantum entanglement of a bipartite quantum state~\cite{bennett}.
As is well known, quantum entanglement can be improved through entanglement purification~\cite{bennett} where a bipartite state
is first transformed to a Werner state  and then two-qubit operations at each sides are needed to improve the quantum entanglement probabilistically.

In this letter, we shall present a theorem (Theorem 2) to maximize the entanglement of a two-qubit mixed
state through single-qubit operations only. The theorem can be used to efficiently
improve the quantum entanglement of a mixed state without the difficult 2-qubit operations.
Explicitly, given a two-qubit mixed state $\rho_{\textrm{in}}=\rho_{12}$, by taking local (non-trace-preserving~\cite{Italy}) maps on qubit 1 and qubit 2 separately, what is the maximally achievable entanglement of the normalized outcome state, and what are the specific maps needed on each qubits.

To make a clear picture of our work  we consider the following example with a pure state $\rho_{\textrm{in}}=|\chi\rangle\langle
\chi|$ and $|\chi\rangle=a|00\rangle + b|11\rangle$ first.
Take the following specific non-trace-preserving map on the first qubit:
\begin{equation}\label{Mfirst}
\varepsilon\otimes I (\rho_{{\rm in}}) = \hat M(\tilde
a,\tilde b)\otimes I\cdot\rho_{{\rm in} }\cdot \hat M(\tilde a,\tilde b)\otimes I,
\end{equation}
where $\hat M(\tilde a,\tilde b)= \tilde a |0\rangle\langle
0|+\tilde b |1\rangle\langle 1|$, and $|\tilde a|^2+|\tilde b|^2=1$.
We have
\begin{equation}\label{m2}
\varepsilon \otimes I (\rho_{{\rm in}}) = \gamma |\chi'\rangle\langle
\chi'|,
\end{equation}
and $\gamma=\sqrt{|a\tilde a|^2+|b\tilde b|^2}$, $|\chi'\rangle =
\frac{a\tilde a}{\gamma}|00\rangle + \frac{b\tilde
b}{\gamma}|11\rangle$. The entanglement concurrence of the outcome
state  is
\begin{equation}
C(|\chi'\rangle\langle\chi'|)= \frac{2|a\tilde ab\tilde b|}{|a\tilde
a|^2+|b\tilde b|^2}.\label{counter}\end{equation} Setting
$|\tilde a|= |b|$ and $|\tilde b|=|a|$, we shall
obtain the maximum output entangled state $|\phi^+\rangle=\frac{1}{\sqrt 2}(|00\rangle+|11\rangle)$  state (up to a normalization factor). Physically, the map $\hat M(\tilde a,\tilde b)$ can be easily realized. For example~\cite{qip}, one can use a
polarization-dependent attenuator, with transmittance proportional to $\tilde a$ for a horizontally polarized photon (state $|0\rangle$) and transmittance
proportional to $\tilde b$ for a vertically polarized photon (state $|1\rangle$). Once we find a photon at the outcome port of the attenuator, the initial
state $|\chi\rangle\langle\chi|$ has been mapped to the outcome state $|\chi'\rangle\langle \chi'|$.

%
\noindent{\em Outline of our work.---}
Our goal is to look for the largest achievable entanglement through local operations, i.e., among all physical maps $\varepsilon\otimes\varepsilon':
\rho_{\textrm{out}}=\varepsilon\otimes\varepsilon'(\rho_{\textrm{in}})$, which map gives out the largest entanglement of the outcome state $\rho_{\textrm{out}}$. Most generally, any local map  $\varepsilon\otimes\varepsilon'$ can be represented in the form of Kruss operators~\cite{Italy}:
\begin{equation}\label{Kruss}
  \rho_{\textrm{out}}=\sum_{i}{\Gamma_{i}\otimes \Gamma_{i}'\cdot \rho_{\textrm{in}}\cdot (\Gamma_{i}\otimes \Gamma_{i}')^{\dagger}}=\sum_i p_i\rho_i,
\end{equation}
where $p_i\rho_i=\Gamma_i\otimes \Gamma_i'\rho_{\textrm{in}}(\Gamma_i\otimes \Gamma_i')^\dagger$.
Denote $C(\rho)$ as the entanglement concurrence~\cite{Wooters} of state $\rho$. Suppose $C(\rho_m)$ is the largest among all $\{C(\rho_i)\}$. Obviously,
\begin{equation}
C(\rho_{\textrm{out}})\;\leq\; \sum_i p_i C(\rho_i)\;\leq\; C(\rho_m).
\end{equation}
Therefore, to find the largest entanglement concurrence of the outcome state among all local maps, we only need to seek it in the following special class of maps:
$\rho_{\textrm{out}} = \mathcal{Q}\otimes \mathcal{Q'}\rho_{\textrm{in}}(\mathcal{Q}\otimes\mathcal{Q'})^{\dagger}$ where $\mathcal Q, \mathcal Q'$ are $2\times 2$ positive matrices. According to the singular-value decomposition, the positive matrix $\mathcal{Q}$($\mathcal{Q'}$) can be decomposed into $\mathcal{Q}=U_{1}DU$($\mathcal{Q'}=U_{1}'D'U'$) where $U_{1},U$($U_{1}',U'$) are unitary matrices and $D$($D'$) is a positive-definite diagonal matrix. Since a unitary transformation plays no role in the entanglement, we only need to consider the positive matrices in the form of $DU$ ($D'U'$).

As shown in Lemma 2, any two-qubit state $\rho_{\textrm{in}}$ can be generated from the maximally-entangled state $|\phi^+\rangle$ acted by a one-sided map $I\otimes \varepsilon'$, i.e.,  $\rho_{\textrm{in}}=I\otimes \varepsilon' (\ket{\phi^{+}}\bra{\phi^{+}})$. Therefore, we can start with entanglement evolution and maximization under non-trace-preserving one-sided maps and then apply the result to the general problem of improving and maximizing quantum entanglement through single-qubit operations.

\noindent{\em Entanglement evolution and maximization under
non-trace-preserving  maps.---}
A non-trace-preserving one-sided map $I\otimes \varepsilon'$ is
fully characterized by $\rho_{\varepsilon'}=I\otimes \varepsilon' (|\phi^+\rangle\langle \phi^+|)$~\cite{liusky}. We
assume
\begin{equation}\label{f}
I\otimes \varepsilon' (|\phi^+\rangle\langle \phi^+|) = f
\rho_{\varepsilon'},\;
I\otimes \varepsilon' (|\psi\rangle\langle \psi|) = f' \rho_{\psi}\;,
\end{equation}
where $f={\rm tr}\left[I\otimes \varepsilon' (|\phi^+\rangle\langle
\phi^+|)\right]$, $f'={\rm tr}\left[I\otimes \varepsilon'
(|\psi\rangle\langle \psi|)\right]$.

A $2\times 2$ pure state
$|\chi\rangle=a|00\rangle +b|11\rangle$ can be rewritten in the form
$|\chi\rangle\langle \chi| = 2\hat M(a,b)\otimes I
(|\phi^+\rangle\langle\phi^+|)\hat M(a,b)\otimes I$. From
Eq.~(\ref{f}) we have $I\otimes
\varepsilon'(|\chi\rangle\langle\chi|)=2f\hat M(a,b)\otimes I
\rho_{\varepsilon'}\hat M(a,b)\otimes I = f' \rho_{\chi}$. We
emphasize here that even though $\rho_{\varepsilon'}$ is normalized,
the operator $2\hat M(a,b)\otimes I \rho_{\varepsilon'}\hat
M(a,b)\otimes I$ is not necessarily normalized.
Define the following function $C$ of an arbitrary
non-negative definite $4\times 4$ matrix (operator) $N$
\begin{equation}\label{con}
C(N)=\rm {max}\{0,\sqrt\xi_1-\sqrt\xi_2-\sqrt\xi_3-\sqrt\xi_4\},
\end{equation}
where $\{\xi_i\}$ are the eigenvalues of $N\cdot \tilde N$, in
descending order,  with $\tilde N=\sigma_y\otimes \sigma_y N^*
\sigma_y\otimes \sigma_y $, and $N^*$ is the complex conjugate of $N$.
If $N$ is a density matrix of a $2\times 2$ system, $C(N)$ is just
the entanglement concurrence of the system~\cite{Wooters}.  With this
definition of $C$, we can summarize the major result, equation (5)
in Ref.~\cite{1} as:
\\{\bf Lemma 1.}
Given any density matrix $\rho_{\varepsilon'}$, if $N = 2\hat
M(a,b)\otimes I\rho_{\varepsilon'} \hat M^\dagger(a,b)\otimes I$,
then
\begin{equation}
C(N)\;=\; C(|\chi\rangle\langle \chi|)\cdot
C(\rho_{\varepsilon'})\;=\;2|ab|C(\rho_{\varepsilon'}).
\end{equation}

However, this is not the entanglement concurrence of $\rho_{\chi}$
because $N$ is not necessarily normalized, even though
$\rho_{\varepsilon'}$ is. Now denote $N=g\rho_\chi$, and $g={\rm
tr}N$. According to the definition of $C$ and $\rho_{\psi}$ in Eq.~(\ref{f}),
\begin{equation}\label{ftr}
C(\rho_{\chi})\ =\ C(N)/g\ =\ 2|ab|C(\rho_{\varepsilon'})/g
\end{equation}
 where $g={\rm tr} N = 2{\rm tr}[\hat M(a,b)\otimes
I\rho_{\varepsilon'}\hat M^\dagger(a,b)\otimes I]$. To avoid
meaningless results, we assume $C(\rho_\varepsilon')>0$ throughout
this paper. Assume that the density matrix of the first qubit of
$\rho_{\varepsilon'}$ is $\rho_0=\rm {tr}_2
\rho_{\varepsilon'}=\left(\begin{array}{cc}c_1 & \alpha
\\ \alpha^* & c_2\end{array}\right)=K_0$, where $\rm{tr}_2$ is the partial
trace over the subspace of the second qubit
 and $c_1 = \langle
0|{\rm tr}_2 \rho_{\varepsilon'}|0\rangle,\; c_2 = \langle 1|{\rm
tr}_2 \rho_{\varepsilon'}|1\rangle.$ Consequently,
\begin{equation}
g =2{\rm tr}[M(a,b)\rho_0M^\dagger(a,b)]=2|a|^2c_1 + 2|b|^2c_2.
\end{equation}
 Therefore, the value of output entanglement
\begin{equation}
C(\rho_\chi)=\frac{2|ab|C(\rho_{\varepsilon'})}{(|a|^2c_1+|b|^2c_2)},
\end{equation}
is maximized when $|a|=\sqrt {c_2},\;|b|=\sqrt {c_1}$, with the maximum value
\begin{equation}
C(\rho_\chi)=\frac{C(\rho_{\varepsilon'})}{2\sqrt{c_1c_2}}.
\end{equation}
More generally, the initial pure state can be
\begin{equation}\label{psi}
|\psi\rangle =I\otimes U |\chi\rangle= \sqrt 2 \hat M(a,b)\otimes U
|\phi\rangle,
\end{equation}
where $U$ is an arbitrary unitary operator.
Given the fact that $U^{*}\otimes U |\phi^+\rangle =|\phi^+\rangle$ for
any unitary $U$, we have
\begin{equation}
 \sqrt 2 \hat M(a,b)\otimes U |\phi^+\rangle =\sqrt 2 \hat
M(a,b)U^T \otimes I|\phi^+\rangle.
\end{equation}
In such a case, we obtain
\begin{equation}
C(\rho_{\psi}) = 2|ab|\cdot C(\rho_{\varepsilon'})/g'
\end{equation}
and $g'={\rm tr}[\hat M(a,b) U^T \otimes
I\rho_{\varepsilon'}U^{*}\hat M(a,b)\otimes I]$. To maximize
$C(\rho_\psi)$, we first fix $U$ and maximize it with $a,b$. Assume
$U^T K_0 U^{*} =\left(\begin{array}{cc}c_1' & \alpha'\\\alpha'^* &
c_2'\end{array}\right).$ The largest value for $C(\rho_\psi)$ is
${C(\rho_{\varepsilon'})}/{2\sqrt{c_1'c_2'}}$, as shown already. To
maximize the value over all $U$, we only need to minimize
$c_1'c_2'$. Since $U$ is unitary, $\det (U^T K_0 U^{*})= \det K_0
$. Therefore $c_1'c_2' =\det K_0 + |\alpha'|^2$, which is minimized
when $\alpha'=0$. Namely,  $C(\rho_{\psi})$ is maximized when
$U^T K_0 U^{*}$ is diagonalized and $\sqrt a =c_2',~ \sqrt b =c_1'$, i.e., $\hat M(a,b) U^T \left({\rm tr}_2\rho_{\varepsilon'}\right)U^{*}\hat M(a,b)= {\rm diag}[1/2,1/2]$. We obtain
\\ {\bf Theorem 1.} Denote $\mathcal Q$ to be a $2\times 2$ positive-definite matrix.
Given the inseparable two-qubit density matrix $\rho_{\textrm{in}}=\rho_{\varepsilon'}= I\otimes \varepsilon' (|\phi^+\rangle\langle\phi^+|)$,
the entanglement of the normalized density matrix $\rho_1 = \mathcal Q \otimes I \rho_{\textrm{in}} (\mathcal Q \otimes I)^\dagger $ maximizes when
$\mathcal Q \left({\rm tr_2}\rho_{\textrm{in}} \right){\mathcal Q}^\dagger = {\rm diag}[1/2,1/2]$ and the
entanglement concurrence
is:
\begin{equation}\label{major}
C_{M}= \frac{C(\rho_{\varepsilon'})}{2\sqrt{\det
\left[{\rm tr}_2\rho_{\varepsilon'}\right]}}.
\end{equation}

\noindent{\em Improving and maximizing quantum entanglement through single-qubit operations.---}
To apply our theorem, we need the following lemma:

{\bf Lemma 2.} Given any $2\times 2$ bipartite mixed state $\rho_{12}$, there exists a map $\varepsilon'$ such that
$\rho_{\textrm{in}}= I\otimes \varepsilon'(|\phi^+\rangle \langle \phi^+|)$.

Note that map $\varepsilon'$ here is in general non-trace-preserving. Since any two-qubit density matrix $\rho_{\textrm{in}}$ can be decomposed into
the mixture of a few pure states, say $\rho_{\textrm{in}}=\sum_i \lambda_i |\psi_i\rangle\langle\psi_i|$. Obviously, for any bipartite pure state $|\psi_i\rangle$, there
always exists a positive operator $\hat M_i'$ such that $|\psi_{i}\rangle = I\otimes \hat M_i' |\phi^+\rangle$. Therefore, we have
$\rho_{\textrm{in}}=\sum_i \lambda_i I\otimes \hat M_i' |\phi^+\rangle \langle \phi^+| I\otimes \hat M_i'^\dagger$. Denoting $I\otimes \varepsilon'(|\phi^+\rangle \langle \phi^+|)= \sum_i  I\otimes \sqrt \lambda_i\hat M_i' |\phi^+\rangle \langle \phi^+| I\otimes \sqrt \lambda_i \hat M_i'^\dagger$ completes the proof.

With Theorem 1 and Lemma 2, we can improve the quantum entanglement of any 2-qubit state $\rho_{\textrm{in}}$ (here and
after, the 2-qibit states are normalized) step by step, with single-qubit operations only.
 Denote $K_{1}={\rm tr}_{2}\rho_{\textrm{in}}$, if
$\det K_{1} < 1/4$, we construct
$\hat{M}_{1}(\tilde{a}_{1}, \tilde{b}_{1})$ and local unitary
$U_{1}$ such that $\hat{M}_{1} U_{1} K_1 U_{1}^{\dagger}
\hat{M}_{1}^{\dagger}={\rm diag[1/2,1/2]} $. The local operation on qubit 1 transforms state $\rho_{\textrm{in}}$ into the outcome state $\rho_{1}=\hat{M}_{1} U_{1}\otimes I \rho_{\textrm{in}} U_{1}^{\dagger}
\hat{M}_{1}^{\dagger}\otimes I$.  According to Theorem 1, the entanglement concurrence
of the outcome state  is
$C(\rho_{1})=C(\rho_{\textrm{in}})/(2\sqrt{\det{K_1}})>C(\rho_{\textrm{in}})$. The normalized density matrix of qubit 1 is $\textrm{tr}_{2}\rho_{1}={\rm diag}[1/2,1/2]$ now, but in general
the normalized density operator of qubit 2 is {\em not} ${\rm diag}[1/2,1/2]$ now.  Using Lemma 2, we know $\rho_1$ can be written in the form of $\rho_1 =\varepsilon \otimes I (|\phi^+\rangle\langle\phi^+|)$. We can now apply Theorem 1 again to further improve the quantum entanglement through operation on qubit 2. Denote $K_1'= {\rm {tr}_{1}}\rho_{1}$.
If $\det(K_1')<1/4$,
we construct new operators $\hat{M}_{1}'$ and $U_{1}'$ such that the density matrix of qubit 2 is ${\rm diag}[1/2,1/2]$ after the operation, i.e., $\hat{M}_{1}' U_{1}' ({\rm tr}_1\rho_{1})
U_{1}^{'\dagger} \hat{M}_{1}^{'\dagger}= {\rm diag}[1/2,1/2]$.
The  operation on qubit 2 leads to a new  outcome state $\rho_{1}'=I\otimes \hat{M}_{1}' U_{1}' \rho_{1} I\otimes
U_{1}^{'\dagger} \hat{M}_{1}^{'\dagger}$.
The operation on qubit 2  improves the entanglement  concurrence to
$C(\rho_{1}')=C(\rho_{1})/(2\sqrt{\det K_1'})
>C(\rho_{1})$.  After the {\em non-trace-preserving} operation above on qubit 2, we have
$K_1'=\textrm{tr}_{1}\rho_{1}'={\rm diag}[1/2,1/2]$, but in general the density matrix of qubit 1 is not ${\rm diag}[1/2,1/2]$, i.e.
$K_2=\textrm{tr}_{2}\rho_{1}'\not= {\rm diag}[1/2,1/2]$ now.  We can
construct new operators $\hat{M}_{2}$ and $U_{2}$ to improve the
entanglement of $\rho_{1}'$. The process will continue step by
step until the determinant of two reduced density matrices are all
equal to $1/4$ after many steps of iterations. Since the entanglement concurrence of a two-qubit state can never be greater than 1 and the entanglement always increases during the iteration process above,
there must exist a limit value of the
entanglement in  the process say, after many steps of iterations, the process gives out the largest entanglement concurrence. This also means that after many steps of iterations, the process always produces a two-qubit state where the reduced density matrices of each qubit are ${\rm diag}[1/2,1/2]$ simultaneously.
Therefore the process that transfers the reduced density matrix of qubit 1 and the reduced density matrix of qubit 2 into ${\rm diag}[1/2,1/2]$ simultaneously always exists and can be written in the following form:
\begin{equation}\label{Improve}
\rho_{f}=\prod_{k=1}^\infty{\hat{M}_{k}U_{k}\otimes\hat{M}_{k}'U_{k}'}\cdot
  \rho_{\textrm{in}} \cdot \prod_{k=1}^\infty{U_{k}^{\dagger}\hat{M}_{k}^{\dagger}\otimes
  U_{k}^{'\dagger}\hat{M}_{k}^{'\dagger}}.
\end{equation}
At the same time, the final state $\rho_{f}$ satisfies the following condition:
\begin{equation}\label{Condf}
  {\rm tr_1}\rho_f={\rm tr_2}\rho_f=I/2.
\end{equation}

\begin{figure}
  \begin{center}
  \includegraphics[width=70mm]{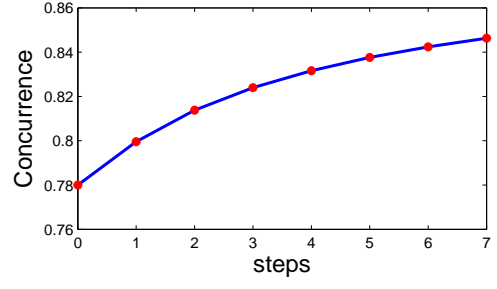}
  \caption{(color online) The concurrence $C$ versus number of iterations of single-qubit operation. \label{fig:fig5}}
  \end{center}
\end{figure}

As an example, consider the imperfect entangled state
\begin{equation}\label{trho}
\tilde{\rho}= 0.1
|\varphi'\rangle\langle\varphi'| +0.12
|\varphi''\rangle\langle\varphi''| + 0.78
|\phi^+\rangle\langle \phi^+|,
\end{equation}
where
$|\varphi'\rangle=\hat{R}(\theta_1)\otimes I|00\rangle$,
$|\varphi''\rangle= \hat{R}(\theta_2)\otimes I|11\rangle$ with
$\theta_1=\pi/5, \theta_2=-3\pi/10$ and $\hat R(\theta) =
\left(
\begin{array}{cc}\cos\theta & \sin\theta \\ \sin\theta & -\cos\theta\
\end{array}\right)$. The entanglement increase through 7 steps of iteration is shown in
Fig.~\ref{fig:fig5}.

The remaining task is to show that, starting from the same state $\rho_{\textrm{in}}$, all final states satisfying Eq.~(\ref{Condf}) have the same value for entanglement concurrence.
\\{\bf Lemma 3.} If  state $\rho_{f}$ satisfies Eq.~(\ref{Condf}), then state $\rho_{f}^{\prime}=U\otimes U^{\prime} \rho_{f} U^{\dagger}\otimes {U^{\prime}}^{\dagger}$ also
satisfies Eq.~(\ref{Condf}). Here $U,U^{\prime}$ are any two unitary operators.

This conclusion is obvious since a unity density operator remains to be unity after any local unitary transformation.

Assume we have two different final states $\rho_{u},\rho_{v}$ obtained
by using different processes from the same initial state, and they satisfy Eq.~(\ref{Condf}). Suppose
$$
\rho_u =\prod_{k=1}^{\infty}{\hat{\mathcal{M}}_{k}\mathcal{U}_{k}\otimes
  \hat{\mathcal{M}}_{k}'\mathcal{U}_{k}'}\cdot
  \rho_{\textrm{in}} \cdot \prod_{k=1}^\infty{\mathcal{U}_{k}^{\dagger}\hat{\mathcal{M}}_{k}^{\dagger}\otimes
  \mathcal{U}_{k}^{'\dagger}\hat{\mathcal{M}}_{k}^{'\dagger}},
$$
$$
\rho_v =\prod_{k=1}^{\infty}{\hat{\mathcal{N}}_{k}\mathcal{V}_{k}\otimes
  \hat{\mathcal{N}}_{k}'\mathcal{V}_{k}'}\cdot
  \rho_{\textrm{in}} \cdot \prod_{k=1}^\infty{\mathcal{V}_{k}^{\dagger}\hat{\mathcal{N}}_{k}^{\dagger}\otimes
  \mathcal{V}_{k}^{'\dagger}\hat{\mathcal{N}}_{k}^{'\dagger}},
$$
where $\hat{\mathcal{M}}_{k},\hat{\mathcal{M}}_{k}',\hat{\mathcal{N}}_{k},\hat{\mathcal{N}}_{k}'$
are projective operators and $\mathcal{U}_{k},\mathcal{U}_{k}',\mathcal{V}_{k},\mathcal{V}_{k}'$
are unitary operators. By using singular-value decomposition, We have
\begin{equation}\label{rhoUV}
  \rho_{v}=\tilde{\mathcal{W}}\mathcal{P}\mathcal{W}\otimes \tilde{\mathcal{W}}^{\prime}
  \mathcal{P}^{\prime}\mathcal{W}^{\prime} \cdot \rho_{u}\cdot \mathcal{W}^{\dagger}\mathcal{P}^{\dagger}
  \tilde{\mathcal{W}}^{\dagger}\otimes {\mathcal{W}^{\prime\dagger}}
  {\mathcal{P}^{\prime}}^{\dagger}{\tilde{\mathcal{W}}^{\prime\dagger}},
\end{equation}
where $\tilde{\mathcal{W}},\mathcal{W},\tilde{\mathcal{W}}^{\prime},\mathcal{W}^{\prime}$ are unitary operators
and $\mathcal{P},\mathcal{P}^{\prime}$ are projective operators defined in Eq.~(\ref{Mfirst}).
Denote $\tilde{\rho}_{w}=\tilde{\mathcal{W}}^{\dagger}\otimes {\tilde{\mathcal{W}}^{\prime\dagger}}
\rho_{v}\tilde{\mathcal{W}}\otimes \tilde{\mathcal{W}}^{\prime}$ and
${\rho}_{w}=\mathcal{W}\otimes \mathcal{W}^{\prime}
\rho_{u}\mathcal{W}^{\dagger}\otimes {\mathcal{W}^{\prime\dagger}}$. We have
\begin{equation}\label{rhoww}
  \tilde{\rho}_{w}=\mathcal{P}\otimes \mathcal{P}^{\prime}\rho_{w}\mathcal{P}^{\dagger}\otimes
  {\mathcal{P}^{\prime\dagger}}.
\end{equation}
According to Lemma 3, we know that $\tilde{\rho}_{w},\rho_{w}$ satisfy Eq.~(\ref{Condf}).
Thus $\mathcal{P}$ and $\mathcal{P}^{\prime}$ must be either identity
or $\textrm{diag}[1,i]$. This indicates that
$\mathcal{P}$ and $\mathcal{P}^{\prime}$ are unitary therefore  the entanglement concurrence of
$\tilde{\rho}_{w}$ and $\rho_{w}$ must be same.
We now obtain the major result of this letter:
\\{\bf Theorem 2.} Given any inseparable two-qubit initial state $\rho_{\textrm{in}}$, the entanglement concurrence can be improved through single-qubit operations provided that the reduced density matrix of any one qubit is not ${\rm diag [1/2,1/2]}$. Among all out-come states $\{\rho_{\textrm{out}}|\rho_{\textrm{out}}=\varepsilon \otimes \varepsilon' (\rho_{\textrm{in}})\}$ through positive-definite local maps, the state $\rho_{\textrm{out}} = \mathcal{Q}\otimes\mathcal{Q'}\rho_{\textrm{in}} \mathcal{Q}^{\dagger}\otimes\mathcal{Q'}^{\dagger}$ has the largest entanglement concurrence
 if the density matrices of each qubit of the outcome state are $I/2$. The corresponding local maps at each side are simply positive-definite matrices $\mathcal{Q},\mathcal{Q'}$ which can be constructed by Eq.~(\ref{Improve}), i.e., $\mathcal{Q}=\prod_{k=1}^\infty{\hat{M}_{k}U_{k}}, \mathcal{Q'}=\prod_{k=1}^\infty{\hat{M}_{k}'U_{k}'}$ where
${\hat{M}_{k}U_{k}}$ ($\hat{M}_{k}'U_{k}'$) diagonalize the state of the first (second) qubit into the form of $I/2$ at the corresponding step. Specifically, $\textrm{tr}_2[(\hat{M}_{k}U_{k}\otimes I) \rho_{k-1}'(\hat{M}_{k}U_{k}\otimes I)^{\dagger}]=I/2$ and $\textrm{tr}_1[(I\otimes\hat{M}_{k}'U_{k}')\rho_{k}(I\otimes \hat{M}_{k}'U_{k}')^{\dagger}]=I/2$ where $\rho_{k}'=\prod_{i=1}^{k}{(\hat{M}_{i}U_{i}\otimes \hat{M}_{i}'U_{i}')\rho_{\textrm{in}}(\hat{M}_{i}U_{i}\otimes \hat{M}_{i}'U_{i}')^{\dagger}}$ and $\rho_{k}=\hat{M}_{k}U_{k}\otimes I \rho_{k-1}'(\hat{M}_{k}U_{k})^{\dagger}\otimes I$.
\\ Remark: In the theorem, we have presented a {\em mathematical} way to construct $\mathcal {Q}, \mathcal{Q'}$ by iteration. We emphasize that, in applying our theorem in a real experiment, one can compute $\mathcal {Q}, \mathcal{Q'}$ and then realize the physical process in {\em only one} step.


\noindent{\em Proposed experiment and numerical simulation.---} We
propose to test Theorem 2 with the initial state $\tilde\rho$ as defined in Eq.~(\ref{trho}).
With many iterations, we have
$\mathcal{Q}=\prod_{k}\hat{M}_{k} U_{k}=
\left(
  \begin{array}{cc}
  0.5875 & -0.8090 \\
  0.0130 &  0.0095
  \end{array}
\right)$ and
$\mathcal{Q'}=\prod_{k}{\hat{M}'}_k U_{k}'=\left(
  \begin{array}{cc}
  0.0106 & -0.0145 \\
  0.8091 &  0.5874
  \end{array}
\right)$.
Then we find that the entanglement concurrence of the final state $\mathcal Q \otimes \mathcal Q' \tilde \rho \mathcal Q^\dagger \otimes {\mathcal Q'}^\dagger$ is 0.8858. Changing matrices $\mathcal Q$ and $\mathcal {Q'}$, the outcome entanglement is always smaller than 0.8858.
Numerical results are presented in Fig.(2) and Fig.(3).
\begin{figure}
  \begin{center}
  \includegraphics[width=70mm]{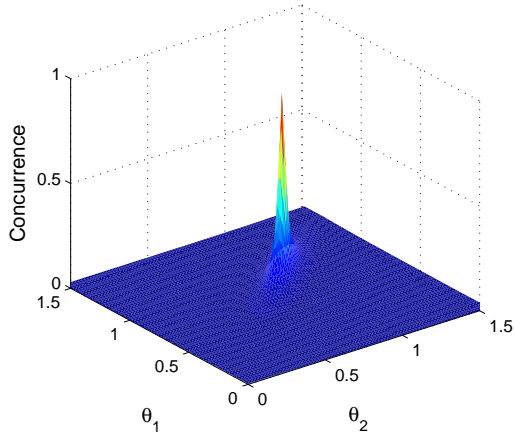}
  \caption{(color online) The concurrence $C$ versus $\theta_1$ and $\theta_2$. Here we have $\mathcal{Q}=\hat{M}(a_1,b_1)U(\theta_1),\mathcal{Q'}=\hat{M}(a_2,b_2)U(\theta_2)$ with $a_1=0.99987,a_2=0.01797,U(\theta)=\hat{R}(\theta)\sigma_z$ and $\sigma_z$ is the Pauli-$z$ matrix. The peak point indicates the maximum concurrence 0.8858 with $\theta_1=0.9427$ and $\theta_2=0.9428$. \label{fig:fig2}
   }
  \end{center}
\end{figure}


\begin{figure}
  \begin{center}
  \includegraphics[width=40mm]{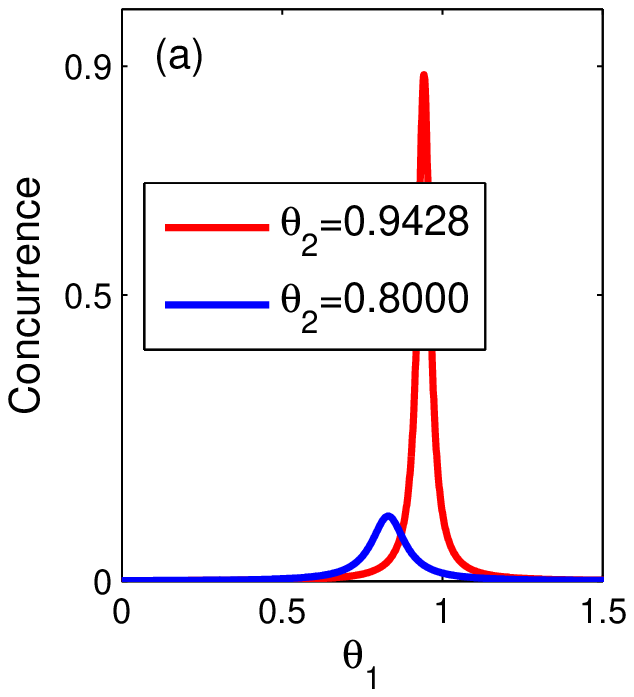}
  \includegraphics[width=40mm]{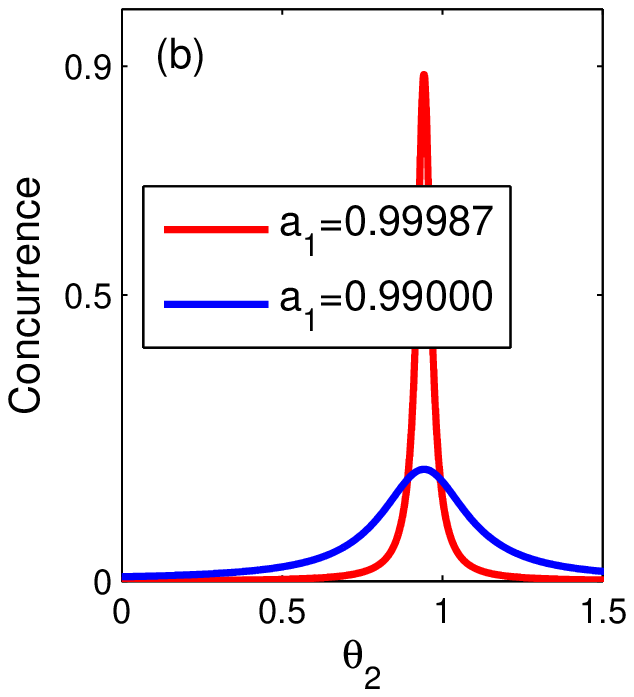}
  \caption{(color online) The concurrence $C$ versus $\theta_1$ [in (a)] and  $\theta_2$ [in
  (b)]. Here we have $\mathcal{Q}=\hat{M}(a_1,b_1)U(\theta_1),\mathcal{Q'}=\hat{M}(a_2,b_2)U(\theta_2)$ where $U(\theta)=\hat{R}(\theta)\sigma_z$. We set $a_1=0.99987,a_2=0.01797$ [in (a)] and $a_2=0.01797,\theta_1=0.9427$ [in (b)]. The peak points indicate the maximum concurrence 0.8858 with $\theta_1=0.9427$ [in (a)] and $\theta_2=0.9428$ [in (b)]. \label{fig:fig3}}
  \end{center}
\end{figure}
{\em Concluding remark.---}
In summary, we have presented explicit results on probabilistically improving and maximizing the quantum entanglement
of a mixed state through single-qubit operations only. Testing schemes are proposed with numerical simulations.
The local operator maximize the outcome entanglement concurrence and can be constructed numerically
by iteration. It is  interesting  to construct
the operators directly from the initial $\rho_{\textrm{in}}$ analytically.

\begin{acknowledgments} XBW is supported by the National Natural
Science Foundation of China under Grant No.~60725416, the National
Fundamental Research Programs of China Grant No. 2007CB807900 and
2007CB807901, and China Hi-Tech Program  Grant No.~2006AA01Z420. FN
acknowledges partial support from the NSA, LPS, ARO, AFOSR, DARPA,  NSF
Grant No.~0726909, JSPS-RFBR Contract No.~09-02-92114, Grant-in-Aid
for Scientific Research (S), MEXT Kakenhi on Quantum Cybernetics,
and the JSPS-FIRST Funding Program.
\end{acknowledgments}

\end{document}